\documentclass[a4paper,USenglish,cleveref, autoref, thm-restate, pdfa]{lipics-v2021}
\usepackage{quiver}
\usepackage{xparse}
\usepackage[numbers,sort]{natbib}
\usepackage{apptools}
\usepackage{longtable}
\usepackage{multicol}
\usepackage{todonotes}
\usepackage{mathtools}

\AtAppendix{\counterwithin{lemma}{section}}
\AtAppendix{\counterwithin{definition}{section}}

\title{Tuple Interpretations and Applications to Higher-Order Runtime Complexity}

\author{Cynthia Kop}
{
    Institute for Computation and Information Sciences,
    Radboud University, The Netherlands \and
    \url{https://www.cs.ru.nl/~cynthiakop}
}
{c.kop@cs.ru.nl}
{https://orcid.org/0000-0002-6337-2544}
{}

\author{Deivid Vale}
{
    Institute for Computation and Information Sciences,
    Radboud University, The Netherlands\and
    \url{https://www.cs.ru.nl/~deividvale}
}
{deividvale@cs.ru.nl}
{https://orcid.org/0000-0003-1350-3478}
{}

\funding{
    The authors are supported by the NWO TOP project ``ICHOR'',
    NWO 612.001.803/7571 and the NWO VIDI project ``CHORPE'',
    NWO VI.Vidi.193.075.
}

\authorrunning{C. Kop and D. Vale}

\Copyright{Cynthia Kop and Deivid Vale}



\hideLIPIcs

\nolinenumbers %

\usepackage{amsmath,amssymb,amsfonts,amscd,amsbsy,mathtools}
\usepackage{microtype}

\usepackage{bussproofs}

\usepackage{xcolor-material}
\usepackage{stmaryrd}

\usepackage{bm}

\newcommand{\sigFont}[1]{\mathsf{#1}}
\newcommand{\sortFont}[1]{\mathsf{#1}}

\newcommand{\compFont}[1]{\mathsf{#1}}

\newcommand{\consFont}[1]{\mathsf{#1}}
\newcommand{\defFont}[1]{\mathtt{#1}}

\newcommand{\metaFont}[1]{\mathtt{#1}}

\newcommand{\nat}{\sortFont{nat}}
\newcommand{\lst}{\sortFont{list}}

\newcommand{\cost}{\compFont{c}}
\newcommand{\size}{\compFont{s}}
\newcommand{\leng}{\compFont{l}}
\newcommand{\mmax}{\compFont{m}}

\newcommand{\zero}{\consFont{0}}
\newcommand{\nil}{\consFont{nil}}
\newcommand{\suc}{\consFont{s}}
\newcommand{\cons}{\consFont{cons}}

\newcommand{\add}{\defFont{add}}
\newcommand{\double}{\defFont{d}}

\newcommand{\map}{\defFont{map}}

\newcommand{\comp}{\defFont{comp}}
\newcommand{\apply}{\defFont{app}}

\newcommand{\asort}{\iota}
\newcommand{\bsort}{\kappa}

\newcommand{\atype}{\sigma}
\newcommand{\btype}{\tau}

\newcommand{\afun}{\sigFont{f}}
\newcommand{\bfun}{\sigFont{g}}

\newcommand{\avar}{x}
\newcommand{\bvar}{y}

\newcommand{\aListVar}{xs}

\newcommand{\aFuncVar}{F}
\newcommand{\bFuncVar}{G}

\newcommand{\aterm}{s}
\newcommand{\bterm}{t}

\newcommand{\num}[1]{\mathsf{#1}}

\newcommand{\costInt}[1]{\mathcal{C}_{#1}}
\newcommand{\costFInt}[1]{\mathcal{F}^\cost_{#1}}
\newcommand{\sizeInt}[1]{\mathcal{S}_{#1}}

\newcommand{\signature}{\mathcal{F}}

\newcommand{\var}{\mathcal{X}}

\newcommand{\bTerms}{T_{\mathcal{B}}(\signature)}
\newcommand{\vars}[1]{\mathtt{vars}(#1)}

\newcommand{\sortset}{\mathcal{B}}
\newcommand{\simpletypeset}{\mathcal{T}_\sortset}

\newcommand{\rules}{\mathcal{R}}
\newcommand{\Nat}{\mathbb{N}}
\newcommand{\NatSizeSet}[1]{\Nat^{\typecount{#1}}}

\newcommand{\arrz}{\to}

\newcommand{\interpret}[1]{\llbracket #1\rrbracket}
\newcommand{\typeinterpret}[1]{\llparenthesis #1 \rrparenthesis}

\newcommand{\typecount}[1]{K[#1]}

\newcommand{\ainterpret}[1]{\llbracket#1\rrbracket_{\alpha, \funcinterpret{}}}
\newcommand{\funcinterpret}[1]{\mathcal{J}_{#1}}

\newcommand{\costGt}{>}
\newcommand{\costGe}{\geq}
\newcommand{\sizeGe}{\sqsupseteq}
\newcommand{\cartGt}{\succ}
\newcommand{\cartGe}{\succcurlyeq}

\newcommand{\pair}[1]{\langle #1 \rangle}

\newcommand{\tuple}[1]{\bm{#1}}
\NewDocumentCommand{\typeVec}{m o}{
    \IfValueTF{#2}{
        {}^{#2}\bm{#1}
    }{
        \bm{#1}
    }
}

\newcommand{\fatlambda}{\lambda\!\!\!\lambda}
\newcommand{\app}{\,}
\newcommand{\semApp}{\cdot}

\newcommand{\unitSet}{\metaFont{unit}}
\newcommand{\unit}{\metaFont{u}}

\newcommand{\dht}[1]{\metaFont{dh}_\rules(#1)}
\newcommand{\rc}[1]{\metaFont{irc}_\rules(#1)}
\newcommand{\ar}{\metaFont{typeOf}}

\newcommand{\termSize}[1]{|#1|}

\newcommand{\arrtype}{\Rightarrow}
\newcommand{\arrfunc}{\longrightarrow}
\newcommand{\arrfuncwm}{\Longrightarrow}

\begin{document}
\maketitle

\begin{abstract}
    Tuple interpretations are a class of algebraic interpretation
    that subsumes both polynomial and matrix interpretations
    as it does not impose simple termination
    and allows non-linear interpretations.
    It was developed in the context of higher-order rewriting
    to study derivational complexity of algebraic functional systems.
    In this short paper, we continue our journey to study the complexity of
    higher-order TRSs by tailoring tuple interpretations to deal with
    innermost
     runtime complexity.
\end{abstract}

\section{Introduction}

The step-by-step computational model induced by term rewriting
naturally gives rise to a \textit{complexity} notion.
Here, complexity is understood as the number of
rewriting steps needed to reach a normal form.
In the rewriting setting,
a \textit{complexity function} bounds the length
of the longest rewrite sequence parametrized by the
size of the starting term.
Two distinct complexity notions are commonly considered:
derivational and runtime.
In the former, the starting term is unrestricted
which allows initial terms with nested function calls.
The latter only considers rewriting sequences
beginning with \textit{basic} terms.
Intuitively,
basic terms are those where a single function call
is performed with \textit{data} objects as arguments.

There are many techniques to bound the runtime complexity of term
rewriting~\cite{nao:moser:08,noschinski:emmes:giesel:13}.
However, most of the literature focuses on the first-order case.
We take a different approach and regard higher-order term rewriting.
We present a technique that takes advantage of tuple
interpretations~\cite{kop:vale:21} tailored to deal with an
innermost rewriting strategy.
The defining characteristic of tuple interpretations
is to allow for a split of the complexity measure
into abstract notions of \emph{cost} and \emph{size}.
The former is usually interpreted as natural numbers,
which accounts for the number of steps needed to reduce terms to normal forms.
Meanwhile, the latter is interpreted as tuples over naturals
carrying abstract notions of size.

\section{Preliminaries}

\subparagraph*{The Syntax of Terms and Rules}
We assume familiarity with the basics of term rewriting.
We will here recall notation for \emph{applicative simply-typed term rewriting systems}.

Let $\sortset$ be a set of base types (or sorts).
The set $\simpletypeset$ of \emph{simple types} is built using the right-associative
$\arrtype$ as follows.
Every $\asort \in \sortset$ is a type of order $0$.
If $\atype, \btype$ are types of order $n$ and $m$ respectively,
then $\atype \arrtype \btype$ is a type of order $\max(n + 1, m)$.
A signature is a non-empty set $\signature$ of function symbols together with a function
$\ar : \signature \arrfunc \simpletypeset$.
Additionally, we assume, for each $\atype \in \simpletypeset$,
a countable infinite set of type-annotated variables $\var_\atype$ disjoint from $\signature$.
We will denote $\afun,\bfun,\dots$ for function symbols and $\avar,\bvar,\dots$ for variables.

This typing scheme imposes a restriction on the formation of terms
which consists of those expressions $\aterm$ such that $\aterm :: \atype$ can be derived
for some type $\atype$ using the following clauses:
(i) $\avar :: \atype$, if $\avar \in \var_\atype$;
(ii) $\afun :: \atype$, if $\ar(\afun) = \atype$;
and (iii) $(\aterm \app \bterm) :: \btype$,
if $\aterm :: \atype \arrtype \btype$ and $\bterm :: \btype$.
Application is left-associative.
We denote $\vars{\aterm}$ for the set of variables occurring in $\aterm$
and say $\aterm$ is \emph{ground}
if $\vars{\aterm} = \emptyset$.
A rewriting rule $\ell \to r$ is a pair of terms of the same type
such that $\ell = \afun \, \ell_1 \dots \ell_m$
and $\vars{\ell} \supseteq \vars{r}$.
An \textit{applicative simply-typed term rewriting system} (shortly denoted TRS), is a set $\rules$ of rules.
The rewrite relation induced by $\rules$ is the smallest monotonic relation that contains $\rules$
and is stable under application of substitution.
A term $\aterm$ is in \textit{normal form} if there is no $\bterm$ such that $\aterm \arrz \bterm$.
The \textit{innermost rewrite relation} induced by $\rules$ is defined as follows:
\begin{itemize}
    \item $\ell\gamma \arrz^i r\gamma$, if $\ell \arrz r \in \rules$
          and all proper subterms of $\ell \gamma$ are in $\rules$-normal form;
    \item $\aterm \app \bterm \arrz^i \aterm' \app \bterm$,
        if $\aterm \arrz^i \aterm'$;
          and $\aterm \app \bterm \arrz^i \aterm \app \bterm'$, if $\bterm \arrz^i \bterm'$.
\end{itemize}
In what follows we only allow for innermost reductions.
So, we drop the \textit{i} from the arrow, and $\aterm \arrz \bterm$ is to be read as $\aterm \arrz^i \bterm$.
We shall use the explicit notation if confusion may arise.

\begin{example}\label{ex:ho-toy-trs}
    We will use a system over the sorts $\nat$ (numbers) and $\lst$ (lists of numbers).
    Let $\zero :: \nat$, $\suc :: \nat \arrtype \nat$,
    $\nil :: \lst$, $\cons :: \nat \arrtype \lst \arrtype \lst$,
    and $F,G \in \var_{\nat \arrtype \nat}$;
    types of other function symbols and variables can be easily deduced.
    \[
    \begin{array}{rclcrcl}
        \map \app \aFuncVar \app \nil & \arrz & \nil & \quad &
        \comp \app \aFuncVar \app \bFuncVar \app \avar & \arrz & \aFuncVar \app (\bFuncVar \app \avar) \\
        \map \app \aFuncVar \app (\cons \app \avar \app \aListVar) & \arrz &
          \cons \app (\aFuncVar \app \avar) \app (\map \app \aFuncVar \app \aListVar) & &
        \apply \app \aFuncVar \app \avar & \arrz & \aFuncVar \app \avar \\
        \double \app \zero & \arrz & \zero & &
        \add \app \avar \app \zero & \arrz & \avar \\
        \double \app (\suc \app \avar) & \arrz & \suc \app (\suc \app (\double \app x)) & &
        \add \app \avar \app (\suc \app \bvar) & \arrz & \suc \app (\add \app \avar \app \bvar) \\
    \end{array}
    \]
\end{example}

\subparagraph*{Functions and orderings}
A quasi-ordered set $(A, \sizeGe)$ consists of a nonempty set $A$
and a quasi-order $\sizeGe$ over $A$.
A well-founded set $(A, \costGt, \costGe)$ is a nonempty set $A$ together with a well-founded order $\costGt$
and a compatible quasi-order $\costGe$ on $A$, i.e.,
$\costGt {\circ} \costGe {\subseteq} \costGt$.
For quasi-ordered sets $A$ and $B$,
we say that a function $f : A \arrfunc B$ is weakly monotonic
if for all $x,y \in A$, $x \sizeGe_A y$ implies $f(x) \sizeGe_B f(y)$.
If $(B, \costGt, \costGe)$ is a well-founded set,
then $\costGt$ and $\costGe$
induce a point-wise comparison on $A \arrfunc B$ as usual.
If $A,B$ are quasi-ordered,
the notation $A \arrfuncwm B$ refers to the set of all
weakly monotonic functions from $A$ to $B$.
Functional equality is extensional.
The unit set is the quasi-ordered set defined by
$\unitSet = (\{\unit\}, \sizeGe)$, where $\unit \sizeGe \unit$.

\section{Higher-Order Tuple Interpretations for Innermost Rewriting}
To define interpretations, we will start by providing an interpretation of \emph{types}
(Def.~\ref{def:cost-size-sets}).
Types $\atype$ are interpreted by tuples $\typeinterpret{\atype}$ that carry information
about cost and size.
We will first show how application works in this newly defined cost-size domain (Def.~\ref{def:semantic-app}).
Interpretation of types will then set the domain for the tuple algebras
we are interested in (Def.~\ref{def:ho-tuple-alg}).

\begin{definition}[Interpretation of Types]\label{def:cost-size-sets}
    For each type $\atype$,
    we define the cost-size tuple interpretation of $\atype$ as
    $\typeinterpret{\atype} = \costInt{\atype} \times \sizeInt{\atype}$
    where $\costInt{\atype}$ (respectively $\sizeInt{\atype}$)
    is defined as follows:
    \begin{align*}
        \costInt{\atype}
        &=
        \Nat \times \costFInt{\atype}
        &
        \sizeInt{\asort}
        & = (\NatSizeSet{\asort}, \sizeGe), \;
        \typecount{\asort} \geq 1 \\
        \costFInt{\asort} & = \unitSet                                                                &
        \sizeInt{\atype \arrtype \btype}  & = \sizeInt{\atype} \arrfuncwm \sizeInt{\btype}                                                              \\
        \costFInt{\atype \arrtype \btype} & = (\costFInt{\atype} \times \sizeInt{\atype}) \arrfuncwm \costInt{\btype},
    \end{align*}
    where $\costFInt{\atype \arrtype \btype}$ ($\sizeInt{\atype \arrtype \btype}$)
    is the set of weakly monotonic functions
    from $\costFInt{\atype} \times \sizeInt{\atype}$ to $\costInt{\btype}$ ($\sizeInt{\atype}$ to $\sizeInt{\btype}$).
    The quasi-ordering on those sets is the induced point-wise comparison.
    The set $\typeinterpret{\atype}$ is ordered as follows:
    $((n,f),s) \cartGt ((m,g),t)$ if $n > m$, $f \costGe g$ and $s \sizeGe t$; and
    $((n,f),s) \cartGe ((m,g),t)$ if $n \geq m$, $f \costGe g$ and $s \sizeGe t$.
\end{definition}

The cost tuple $\costInt{\atype} = \Nat \times \costFInt{\atype}$ of $\typeinterpret{\atype}$
holds the cost information of reducing a term of type $\atype$
to its normal form.
It is composed of a numeric and functional component.
Base types, which are naturally not functional,
have the unit set for $\costFInt{\asort}$;
the cost tuple of a base type is then $\costInt{\asort} = \Nat \times \unitSet$.
Functional types do possess an intrinsically functional component
(the cost of \emph{applying} the function),
which in our setting is expressed by
$\costFInt{\atype \arrtype \btype} = \costFInt{\atype} \times \sizeInt{\atype} \arrfuncwm \costInt{\btype}$.
For functional types the numeric component represents the cost of partial application.

To determine the number $\typecount{\asort}$,
associated to each sort $\asort$,
we use a semantic approach
that takes the intuitive meaning of the sort
we are interpreting into account.
The sort $\nat$ for instance represents natural numbers,
which we implement in unary format.
Hence,
it makes sense to reckon the number of successor symbols
occurring in terms of the form $(\suc^n \app \zero) :: \nat$
as their \textit{size}.
This gives us $\typecount{\nat} = 1$.
Another example is the sort $\lst$ (of natural numbers):
it is natural to regard measures like
\textit{length} and \textit{maximum element size}.
This results in $\typecount{\lst} = 2$.
Example \ref{ex:int-constructors} below shows
how to interpret data constructors using this intuition.

The next lemma expresses the soundness of our approach,
that is,
cost-size tuples define a well-founded domain for the interpretation of types.

\begin{lemma}\label{lemma:cs-sets-cs-tuple}
    For each type $\atype$,
    the set $\costInt{\atype}$ is well-founded and $\sizeInt{\atype}$ quasi-ordered.
    Their product, that is, $(\typeinterpret{\atype}, \cartGt, \cartGe)$,
    is well-founded.
\end{lemma}

\subparagraph*{Semantic Application}
To interpret each term $\aterm :: \atype$ to an element of $\typeinterpret{\atype}$
(Def.~\ref{def:ho-tuple-alg}), we will need a notion of application for cost-size tuples.
Specifically, given a functional type $\atype \arrtype \btype$,
a cost-size tuple $\tuple{f} \in \typeinterpret{\atype \arrtype \btype}$,
and $\tuple{x} \in \typeinterpret{\atype}$,
our goal is to define the application $\tuple{f} \semApp \tuple{x}$ of $\tuple{f}$ to $\tuple{x}$.
Let us illustrate the idea with a concrete example:
consider the type $\atype = (\nat \arrtype \nat) \arrtype \lst \arrtype \lst$,
which is the type of $\map$ defined in Example \ref{ex:ho-toy-trs}.
The function $\map$ takes as argument a function $F$ of type $\nat \arrtype \nat$
and a list $q$, and applies $F$ to each element of $q$.
The cost interpretation of $\map$ is a functional in $\costInt{\atype}$ parametrized by
functional arguments carrying the cost and size information of $F$
and a cost-size tuple for $q$.
\[
    \Nat \times
    \overbrace{
    (
        \underbrace{
        (\unitSet \times \Nat \arrfuncwm \Nat \times \unitSet)
        }_{\text{cost of } F}
        \times
        \underbrace{
        (\Nat \arrfuncwm \Nat)
        }_{\text{size of } F}
        \arrfuncwm
        (
            \Nat
            \times
            (
            \underbrace{
                \unitSet
            }_{\text{cost of } q} \times
            \underbrace{
                \Nat^2
            }_{\text{size of } q}
            \arrfuncwm
            \Nat \times \unitSet
            )
        )
    )
    }^{\text{the functional cost of } \map},
\]
Hence,
we write an element of such space as the tuple $(n, f^\cost)$.
Size sets are somewhat simpler with
$\underbrace{(\Nat \arrfuncwm \Nat)}_{\text{size of } F}
\arrfuncwm \underbrace{\Nat^2}_{\text{size of } q} \arrfuncwm \Nat^2$.
Therefore, a functional cost-size tuple $\tuple{f}$
is represented by $\tuple{f} = \pair{(n, f^\cost), f^\size}$.
An argument to such a cost-size tuple
is then an element in the domain of $f^\cost$ and $f^\size$, respectively.
Therefore, we apply $\tuple{f}$ to a cost-size tuple $\tuple{x}$
of the form $\pair{(m, g^\cost), g^\size}$ where
$g^\cost$ is the cost of computing $F$ and $g^\size$
is the size of $F$.
We proceed by applying the respective functions,
so $f^\cost(g^\cost, g^\size) = (k, h)$ belongs to $\costInt{\lst}$,
and add the numeric components together obtaining:
$\tuple{f} \semApp \tuple{x} =
\pair{(n + m + k, f^\cost(g^\cost, g^\size)), f^\size(g^\size)}$.
Notice that this gives us a new cost-size tuple with cost component in
$\Nat \times (\costInt{\lst} \arrfuncwm \costInt{\lst})$
and size component in $\sizeInt{\lst} \arrfuncwm \sizeInt{\lst}$.

\begin{definition}\label{def:semantic-app}
    Let $\atype \arrtype \btype$ be an arrow type,
    $\tuple{f} = \pair{(n, f^\cost),f^\size} \in \typeinterpret{\atype \arrtype \btype}$,
    and $\tuple{x} = \pair{(m, g^\cost), g^\size} \in \typeinterpret{\atype}$.
    The application of $\tuple{f}$ to $\tuple{x}$,
    denoted $\tuple{f} \semApp \tuple{x}$, is defined by:
    \begin{center}
        let $f^\cost(g^\cost, g^\size) = (k,h)$; then
        \( \pair{(n, f^\cost), f^\size} \semApp \pair{(m, g^\cost), g^\size} =
        \pair{(n + m + k, h), f^\size(g^\size)}
        \)
    \end{center}

\end{definition}

Semantic application is left-associative
and respects a form of application rule.
\begin{lemma}\label{lemma:semApp-type-sound}
    If $\tuple{f}$ is in $\typeinterpret{\atype \arrtype \btype}$
    and $\tuple{x}$ is in $\typeinterpret{\atype}$,
    then $\tuple{f} \semApp \tuple{x}$ belongs to $\typeinterpret{\btype}$.
\end{lemma}

\begin{remark}
    In order to ease notation,
    we project sets $\pi_1 : A \times \unitSet \arrfunc A$
    and $\pi_2 : \unitSet \times A \arrfunc A$
    and compose functions with projections,
    so a function in $\unitSet \times A \arrfuncwm B \times \unitSet$
    is lifted to a function in $A \arrfuncwm B$.
    The functional cost of $\map$ is then read as follows:
    \begin{equation*}
        \Nat \times
        \overbrace{
        (
            \underbrace{
            (\Nat \arrfuncwm \Nat)
            }_{\text{cost of } F}
            \times
            \underbrace{
            (\Nat \arrfuncwm \Nat)
            }_{\text{size of } F}
            \arrfuncwm
            (
                \Nat
                \times
                (
                \underbrace{
                    \Nat^2
                }_{\text{size of } q}
                \arrfuncwm
                \Nat
                )
            )
        )
        }^{\text{the functional cost of } \map}
    \end{equation*}
    The $\Nat$ component of $C_{\atype \arrtype \btype}$ is specific to innermost rewriting
    (it does not occur in \cite{kop:vale:21}).
    We need this to handle rules of non-base type; for
    example, if $\add \app \zero \to \defFont{id}$, then the cost tuple of
    $\add\ \zero$ is $(1,\fatlambda x.0)$.
    However, since in \emph{most} cases the first component is $0$, we will typically omit
    these zeroes and simply write for instance
    $\fatlambda F q. f^\cost(F, q)$ instead of
    $(0, \fatlambda F. \pair{0, \fatlambda q. f^\cost(F, q)})$.
    To compute using Definition \ref{def:semantic-app} we still use the complete form.
\end{remark}

Tuple algebras are higher-order weakly monotonic algebras~\cite{fuh:kop:12}
with cost-size tuples as interpretation domain.

\begin{definition}[Higher-order tuple algebra]\label{def:ho-tuple-alg}
    A higher-order tuple algebra over a signature $(\sortset, \signature, \ar)$
    consists of:
    (i) a family of cost/size tuples $\{ \typeinterpret{\atype} \}_{\atype \in \simpletypeset}$
    and
    (ii) an interpretation function $\funcinterpret{}$
    which maps each $\afun \in \signature$ of type $\atype$
    to a cost-size tuple in $\typeinterpret{\atype}$.
\end{definition}

\begin{example}\label{ex:int-constructors}
    Following the semantics discussed previously,
    we interpret the constructors for both $\nat$ and $\lst$ as follows.
    We call the first component of $\sizeInt{\lst}$
    \textit{length} and the second \textit{maximum element size}.
    Those are abbreviated using the letters $\leng$ and $\mmax$, respectively.
    \begin{align*}
        \funcinterpret{\zero} & = \pair{0, 0} &
        \funcinterpret{\suc}  & = \pair{
            \fatlambda x. 0,
            \fatlambda x. x + 1
        }\\
        \funcinterpret{\nil}  & = \pair{0, \pair{0,0}} &
        \funcinterpret{\cons} & = \pair{
            \fatlambda x q. 0,
            \fatlambda x q. \pair{q_\leng + 1, \max(x, q_\mmax)}
        }
    \end{align*}

    The cost-size tuples for $\zero$ and $\nil$ are all $0$s, as expected.
    The size components for $\suc$ and $\cons$ describe
    the increase in size when new data is created.
    We interpret functions from Example \ref{ex:ho-toy-trs} as follows:
    \[
    \begin{array}{rclrcl}
        \funcinterpret{\apply} & = &
        \pair{
            \fatlambda F x. F^\cost(x) + 1,
            \fatlambda F x. F^\size(x)
        }\\
        \funcinterpret{\double} & = &
        \pair{
            \fatlambda x. x + 1,
            \fatlambda x. 2x
        }\\
        \funcinterpret{\add} & = &
        \pair{
            \fatlambda x y.
                y + 1,
            \fatlambda x y. x + y
        }\\
        \funcinterpret{\comp} & = &
        \pair{
            \fatlambda F G x. F^\cost(G^\size(x_\size)) + 1,
            \fatlambda F G x. F^\size(G^\size(x))
        }\\
        \funcinterpret{\map} & = &
        \pair{
            \fatlambda Fq. q_\leng F^\cost(q_\mmax) + 1,
            \fatlambda F q.\pair{q_\leng, F^\size(q_\mmax)}
        }
    \end{array}
    \]
\end{example}

A valuation $\alpha$ is a function that maps each $\avar :: \atype$
to a cost-size tuple in $\typeinterpret{\atype}$.
Due to innermost strategy,
we can assume the interpretation of every variable
$\avar :: \asort$ has zero cost.
This is formalized by assigning
$\alpha(x) = \pair{(0, \unit), x^s}$,
for all $\avar \in \var$ of base type.
In this paper,
we shall only consider valuations that satisfy this property.
Variables of functional type, however,
may carry cost information even though
any instance of a redex needs to be normalized.
Hence, we set $\alpha(F) = \pair{(0, f^\cost), f^\size}$
when $F :: \atype \arrtype \btype$.

\begin{definition}\label{def:term-int}
    We extend $\funcinterpret{}$ to an interpretation
    $\interpret{\cdot}_{\alpha, \funcinterpret{}}$
    of terms as follows:
    \begin{align*}
        \ainterpret{\avar} & = \alpha(\avar)
        &
        \ainterpret{\afun}
         & = \pair{
             (n, \funcinterpret{\afun}^\cost),
             \funcinterpret{\afun}^\size
         }, n \in \Nat
        &
        \ainterpret{\aterm \app \bterm}
        & =
        \ainterpret{\aterm} \semApp \ainterpret{\bterm}
    \end{align*}
\end{definition}
We are interested in interpretations satisfying a compatibility requirement:
\begin{theorem}[Innermost Compatibility Theorem]\label{theorem:compatibility}
    Let $\alpha$ be a valuation.
    If $\ainterpret{\ell} \cartGt \ainterpret{r}$
    for all rules $\ell \to r \in \rules$,
    then
    $\ainterpret{\aterm} \cartGt \ainterpret{\bterm}$, whenever $\aterm \arrz_\rules^i \bterm$.
\end{theorem}

One can check that the TRS from Example \ref{ex:ho-toy-trs} interpreted as in Example \ref{ex:int-constructors} satisfy the compatibility requirement.

\section{Higher-Order Innermost Runtime Complexity}

In this section,
we briefly limn how the cost-size tuple machinery
allow us to reason about innermost runtime complexity.
We start by reviewing basic definitions.

\begin{definition}
    A symbol $\afun \in \signature$ is a \textit{defined symbol}
    if it occurs at the head of a rule, i.e.,
    there is a rule $\afun \app \ell_1 \app \dots \app \ell_k \to r \in \rules$.
    A symbol $\consFont{c}$ of order at most 1
    is a \textit{data constructor} if it is not a defined symbol.
    A \textit{data term} has the form $\consFont{c} \app d_1 \app \dots \app d_k$
    with $\consFont{c}$ a constructor and each $d_i$ a data term.
    A term $\aterm$ is \textit{basic} if $\aterm :: \asort$ and
    $\aterm$ is of the form $\afun \app d_1 \app \dots \app d_m$
    with $\afun$ a defined symbol and all $d_1, \dots, d_m$ data terms.
    The set $\bTerms$ collects all basic terms.
\end{definition}

\begin{remark}
    Notice that our notion of data is intrinsically first-order.
    This is motivated by applications of rewriting to full program analysis
    where even if higher-order functions are used a program has type
    $\asort_1 \arrtype \dots \arrtype \asort_m \arrtype \bsort$.
    The sorts $\asort_i$ are the input data types
    and $\bsort$ the output type of the program.
\end{remark}

\begin{definition}
    The innermost derivation height of $\aterm$ is
    $\dht{\aterm} = \{ n \mid \exists \bterm : \aterm \arrz^n \bterm \}$.
    The innermost runtime complexity function with respect to a TRS $\rules$ is
    $\rc{n} =
    \max \{ \dht{\aterm} \mid
        \aterm \in \bTerms \wedge \termSize{\aterm} \leq n
    \}$.
\end{definition}

To reasonably bound the innermost runtime complexity of a TRS $\rules$,
we require that size interpretations of constructors have their components
bounded by an additive polynomial, that is,
a polynomial of the form
$\fatlambda x_1 \dots x_k. \sum_{i = 1}^k x_i + a$, with $a \in \Nat$.

We can build programs by adding a new $\defFont{main}$ function
taking data variables as arguments
and combine it with rules computing functions, including higher-order ones.
For instance, using rules from Example \ref{ex:ho-toy-trs},
we can compute a program that adds a number $x$ to every element in a list $q$ as follows:
$\defFont{main} \app x \app q \to \map \app (\add \app x) \app q$.
Hence, computing this program on inputs $\num{n}$ and list $q$
is equivalent to reducing the term $\defFont{main} \app \num{n} \app q$ to normal form.
Its runtime complexity is therefore bounded by the cost-tuple of
$\interpret{\defFont{main} \app \num{n} \app q}$.

\section{Conclusion}
In this short paper, we shed light
on how to use cost-size tuple interpretations
to bound innermost runtime complexity of higher-order systems.
We defined a new domain of interpretations
that takes the intricacies of innermost rewriting into account
and defined how application works in this setting.
The compatibility result allows us
to make use of interpretations as a way
to bound the length of derivation chains,
as it is expected from an interpretation method.
As current, and future work, we are working on automation techniques
to find interpretations and develop a completely rewriting-based automated tool
for complexity analysis of functional programs.

\bibliography{references}
\end{document}